\documentclass[twocolumn,10pt,letterpaper]{quantumarticle}
\pdfoutput=1
\usepackage[numbers]{natbib}
\usepackage[utf8]{inputenc}
\usepackage[T1]{fontenc}
\usepackage{amsmath,amssymb}
\usepackage{graphicx}
\usepackage[colorlinks=true,allcolors=quantumviolet]{hyperref}
\usepackage{xurl}
\usepackage{braket}

\begin{document}

\title{When does dissipation help neural surrogates learn open quantum dynamics?}

\author{Alauddin Ahmed}
\affiliation{Mechanical Engineering, University of Michigan, Ann Arbor, MI 48109, USA}
\email{alauddin@umich.edu}
\orcid{0000-0003-4238-1561}

\maketitle

\begin{abstract}
Dissipation is usually viewed as an obstacle to predicting quantum dynamics, yet it can also contract trajectories toward steady states and thereby suppress accumulated prediction errors, leaving it unclear whether dissipation ultimately helps or hinders the learnability of open quantum dynamics. We investigate this question using Neural Ordinary Differential Equation (NODE) surrogates for open Heisenberg XYZ spin chains. Closed-system learnability deteriorates rapidly with system size, culminating in a static-prediction collapse at four qubits; dissipation reverses this trend, creating a broad high-fidelity regime at intermediate system sizes, while at four qubits a fidelity-aware objective recovers learnable rollout structure that is absent under closed-system training. Comparison against static and steady-state baselines reveals that dissipation improves performance through two fundamentally different mechanisms: at weak-to-moderate dissipation the surrogate captures nontrivial transient dynamics and substantially outperforms trivial predictors, whereas at stronger damping high fidelity increasingly reflects trajectory simplification toward the steady state rather than improved learned dynamics. These results show that dissipation can enhance the learnability of open quantum dynamics, but that fidelity alone is insufficient to distinguish genuine dynamical learning from steady-state trivialization: dissipative contraction and trajectory simplification are distinct effects that peak in different regimes and should be disentangled when evaluating learned quantum-dynamical surrogates.
\end{abstract}

\section{Introduction}

Simulating the dynamics of open quantum systems is a recurring
computational bottleneck across quantum information, condensed matter,
and quantum chemistry. The standard equation of motion is the
Gorini--Kossakowski--Sudarshan--Lindblad (GKSL) master
equation,\cite{gorini1976,lindblad1976,breuer2007,manzano2020} whose direct integration scales
exponentially in the number of qubits and rapidly exhausts the reach of
conventional methods. This bottleneck has motivated sustained interest
in learned surrogate models, in which a neural network is trained to
predict the time evolution of the system from a modest number of
reference trajectories.\cite{schollwock2011,verstraete2004}

Among recent proposals, Neural Ordinary Differential Equation (NODE)
surrogates\cite{chen2018node,kidger2022,rubanova2019} are particularly attractive. They
parameterize the generator of the dynamics directly as a learned vector
field, are continuous in time by construction, and admit
memory-efficient training via the adjoint sensitivity method. NODEs have
been applied to closed and open quantum dynamics in several recent
studies,\cite{choi2022,chen2022} establishing the feasibility of the
approach in representative settings.

In practice, however, the empirical behavior of NODE surrogates on open
quantum systems remains poorly characterized. The wider
neural-network-for-quantum-many-body literature has overwhelmingly
focused on closed-system wavefunction representations using neural
network quantum states,\cite{carleo2017,torlai2018tomography,carleo2019netket} with open-system
extensions building primarily upon neural density
operators\cite{hartmann2019,nagy2019,vicentini2019,yoshioka2019} and time-dependent variational
approaches.\cite{reh2021}

Two fundamental questions remain comparatively underexplored. Given the
central role of dissipation in noisy intermediate-scale quantum (NISQ)
devices,\cite{preskill2018,bharti2022} these questions warrant direct empirical
study. The first is how learnability scales with system size for open
quantum dynamics: does NODE surrogate fidelity degrade gracefully with
qubit count, or does it collapse abruptly? The second is how dissipation
strength shapes that learnability: does dissipation, which is ubiquitous
in realistic quantum hardware, help or hinder a
surrogate\textquotesingle s ability to predict the trajectory? These
questions matter because the regime of practical interest, simulating
NISQ devices, is precisely the regime of open dynamics at qubit counts
beyond textbook examples.

There are a priori reasons to expect dissipation to matter for
learnability. Closed quantum dynamics preserve coherent oscillations
indefinitely, with no contraction of trajectory errors. Consequently,
rollout errors in a learned vector field can accumulate over the
integration horizon. This challenge is expected to become more severe as
the state-space dimension grows. Dissipation, by contrast, contracts
trajectories toward steady states, providing a natural mechanism by
which learned errors are damped rather than amplified. This intuition
has a concrete parallel in the broader neural ordinary differential
equation literature, where contractive vector fields are known to yield
more robust and accurate learned trajectories than non-contractive ones.
Explicit architectural enforcement of contractivity has been shown to
improve both stability and generalization.\cite{zakwan2023} At the
same time, strong dissipation may erase the structure the surrogate is
meant to capture, suggesting that the optimal regime for learning may be
intermediate rather than extreme. Despite this intuition, controlled
empirical sweeps over dissipation strength and system size remain
limited for open quantum surrogates of NISQ systems.

A second piece of structure motivates the surrogate design. The natural
training loss for a density-operator surrogate is a matrix-norm error,
specifically the Frobenius norm or an element-wise state mean squared
error (MSE), but the physically meaningful benchmark for quantum
trajectories is Uhlmann fidelity. These metrics need not align as
Hilbert-space dimension grows, because a small matrix-norm error does
not necessarily imply high fidelity, as our d = 16 results illustrate in
Section 5. We do not claim a general scaling law relating Frobenius
error and fidelity; rather, we observe a severe mismatch in the present
d = 16 setting, which motivates the fidelity-aware training objective
introduced in Section 3.

We address these questions with a controlled empirical study. We train
NODE surrogates on an open Heisenberg XYZ chain of n \(\in\) \{2, 3, 4\}
qubits governed by a nearest-neighbor Heisenberg XYZ Hamiltonian with
local dephasing and amplitude-damping channels. We sweep the
dimensionless dissipation strength \(\gamma\)/J across four orders of magnitude
(\(\gamma\)/J \(\in\) \{0, 0.01, 0.1, 1, 10\}, with an additional bounding point at \(\gamma\)/J
= 100) and the training-set size D \(\in\) \{500, 2000, 8000\}. Dissipative
XYZ and anisotropic-Heisenberg models of the same class we study here
have previously served as benchmark problems for neural-network
approaches to open quantum systems, though that work targeted the
nonequilibrium steady state through variational Monte Carlo rather than
the full transient dynamics learned here.\cite{hartmann2019,nagy2019,vicentini2019}
Reference dynamics in this work are obtained using the QuTiP
library.\cite{qutip1,qutip2,qutip5} All hyperparameters other than (n, \(\gamma\), D)
are held fixed, and three independent random seeds are used per
parameter configuration to quantify seed variability.

We report trajectory-averaged fidelity, an early-time and late-time
window decomposition of the rollout, and physical-state-validity metrics,
namely trace deviation and the minimum eigenvalue of the predicted
density operator.

Our findings are threefold. First, closed-system fidelity degrades
markedly with qubit count, with mean trajectory fidelity dropping from
$\bar{F}_{\mathrm{avg}}$ \(\approx\) 0.97 for n = 2 to 0.50 for n = 3 and 0.29 for n = 4. This decline
is essentially insensitive to a 16-fold increase in training data,
consistent with a representational rather than a data-limited failure
mode. Second, at n = 3, weak-to-moderate dissipation substantially
improves trajectory-averaged fidelity over the closed-system baseline,
yielding a broad sweet spot near \(\gamma\)/J \(\in\) \{0.1, 1.0\} and a slight
degradation at \(\gamma\)/J = 10. The improvement is concentrated in the
late-time window of the rollout, where $\bar{F}_{\mathrm{late}} > \bar{F}_{\mathrm{early}}$,
consistent with dissipative contraction of trajectory errors.

Third, at n = 4 with naive state mean squared error (MSE) training,
surrogates converge to a near-static prediction whose matrix error is
small but whose fidelity remains near that of a trivial static
predictor. We attribute this behavior to the Frobenius--fidelity
mismatch discussed above. In the present d = 16 setting, this mismatch
renders a static prediction a spurious low-loss minimum from which the
optimizer does not escape. Introducing a differentiable trace-distance
loss term recovers learnability across three random seeds and restores
the $\bar{F}_{\mathrm{late}} > \bar{F}_{\mathrm{early}}$ signature observed at n = 3. This
recovery comes at a measurable cost to the physical admissibility of the
predicted states, motivating future work on completely positive
trace-preserving (CPTP)-constrained architectures.\cite{vicentini2022}

We interpret all results within the tested architecture, training
horizon, and training budgets; detailed limitations are discussed in
Section 6. Section 2 describes the physical model and surrogate
architecture. Section 3 outlines the controlled experimental design.
Section 4 presents the n = 2 and n = 3 results. Section 5 analyzes the n
= 4 optimization pathology and its fidelity-aware resolution. Section 6
concludes with a discussion of implications, limitations, and outlook.

\section{Model: open Heisenberg XYZ chain with local Lindblad dissipators}

\subsection{Hamiltonian}

We study an open quantum spin chain of \(n \in \{ 2,3,4\}\) qubits with
nearest-neighbor Heisenberg XYZ interactions and open boundary
conditions.\cite{mikeska2004} The coherent part of the dynamics is
generated by the Hamiltonian

\begin{equation}
H = \sum_{i = 1}^{n - 1}\left( J_{x}\sigma_{i}^{x}\sigma_{i + 1}^{x} + J_{y}\sigma_{i}^{y}\sigma_{i + 1}^{y} + J_{z}\sigma_{i}^{z}\sigma_{i + 1}^{z} \right)
\label{eq:1}
\end{equation}

where \(\sigma_{i}^{\alpha}\) (\(\alpha \in \{ x,y,z\}\)) denotes the
Pauli operator acting on site \(i\). The system is defined on the
\(n\)-qubit Hilbert space
\(\mathcal{H} = \left( \mathbb{C}^{2} \right)^{\otimes n}\) of dimension
\(d = 2^{n}\).

Throughout this work, we fix the coupling constants to
\(J_{x} = J_{y} = 1\) and \(J_{z} = 0.8\), corresponding to an
anisotropy parameter \(\Delta = J_{z}/J_{x} = 0.8\), placing the system
away from the isotropic Heisenberg point (\(\Delta = 1\)). We use units
in which \(J \equiv J_{x} = J_{y} = 1\), so that all energies,
dissipation rates, and times are dimensionless, with time measured
implicitly in units of \(J^{- 1}\).

\subsection{Dissipative Dynamics}

The full state evolution follows the
Gorini--Kossakowski--Sudarshan--Lindblad (GKSL) master equation

\begin{equation}
\frac{d\rho}{dt} = - i[ H,\rho] + \sum_{k}^{}\gamma_{k}\left( L_{k}\rho L_{k}^{\dagger} - \frac{1}{2}\{ L_{k}^{\dagger}L_{k},\rho\} \right)
\label{eq:2}
\end{equation}

where the index \(k\) runs over both jump-operator families and all sites
\(i \in \{ 1,\ldots,n\}\):

The first family is local dephasing, \(L_{\phi,i} = \sigma_{i}^{z}\) with
rate \(\gamma_{\phi}\), which preserves populations in the \(z\)-basis,
conserving the local expectation value \(\langle\sigma_{i}^{z}\rangle\),
while destroying the coherence between \(\lvert 0\rangle\) and
\(\lvert 1\rangle\) on site \(i\). The second family is local amplitude
damping, \(L_{\downarrow ,i} = \sigma_{i}^{-}\) with rate
\(\gamma_{\downarrow}\), where
\(\sigma_{i}^{-} = (\sigma_{i}^{x} - i\sigma_{i}^{y})/2\). We adopt the
convention \(\sigma^{z}\lvert 0\rangle = -\lvert 0\rangle\), so that
\(\lvert 0\rangle\) is the lower state and this dissipator drives each
site toward \(\lvert 0\rangle\). Because each \(\sigma_{i}^{-}\)
annihilates \(\lvert 0\rangle\) and each \(\sigma_{i}^{z}\) leaves
\(\lvert 0\ldots 0\rangle\langle 0\ldots 0\rvert\) invariant, the fully
polarized product state
\(\lvert 0\ldots 0\rangle\langle 0\ldots 0\rvert\) is a dark state of the
dissipators and serves as a steady state of the dynamics.

Throughout the controlled sweep we set
\(\gamma_{\phi} = \gamma_{\downarrow} \equiv \gamma\) and vary the single
dimensionless dissipation strength \(\gamma/J\) across four orders of
magnitude (\(\gamma/J \in \{ 0,0.01,0.1,1,10\}\), plus the bounding
point \(\gamma/J = 100\); see \S3). Holding
\(\gamma_{\phi} = \gamma_{\downarrow}\) removes a confounding
two-parameter sweep and keeps the controlled study tractable. The
trade-off is that we cannot disentangle the individual contributions of
dephasing and amplitude damping, a restriction we revisit in \S6.

\subsection{Initial States and Trajectory Observables}\label{sec:initial}

The test dataset is drawn from a stratified ensemble comprising 80\%
Haar-random pure states (sampled as \(U \mid 0\ldots 0\rangle\) with
\(U\) a Haar-random unitary on \(\mathcal{H}\)), 10\% computational-basis
states sampled uniformly from
\(\left\{ \mid 0\ldots 0\rangle, \mid 0\ldots 01\rangle,\ldots, \mid 1\ldots 1\rangle \right\}\),
and 10\% Bell-state initial conditions on the first two qubits, with the
remaining \(n - 2\) qubits initialized in \(\mid 0\rangle\). This mixed
ensemble probes generalization beyond the Haar-distributed training
ensemble while sampling a range of entanglement structures. The 80/10/10
split was chosen heuristically before training and is held fixed across
all parameter configurations. Training trajectories are drawn
exclusively from the Haar ensemble, as discussed in Section 6.

For each parameter configuration, we generate trajectories \(\rho(t)\) on
a uniform time grid spanning \(t \in [0, T]\) with total duration
\(T = 10/J\), sampled at 200 grid points inclusive of both endpoints
(spacing \(\Delta t = T/199 \approx 0.0503/J\)). The integration horizon was chosen to allow dissipative
trajectories to approach steady state while capturing several coherent
oscillations in the closed-system regime. Reference trajectories are
obtained by direct integration of the GKSL master equation using the
QuTiP 5 library with adaptive numerical ODE solvers, with relative and
absolute error tolerances set to
\(\text{rtol} = \text{atol} = 1 \times 10^{- 8}\) across all dissipation
strengths.

\subsection{Fidelity metrics}\label{sec:fidelity}

The headline performance metric is the trajectory-averaged Uhlmann
fidelity between predicted and true density
matrices,\cite{uhlmann1976,jozsa1994,nielsen2010}

\begin{align}
\label{eq:fidavg}
\bar{F}_{\mathrm{avg}} &= \langle F(\rho_{\mathrm{pred}}(t),\rho_{\mathrm{true}}(t))\rangle_{t,\mathrm{test}}, \\
F(\rho,\sigma) &= \left(\mathrm{Tr}\sqrt{\sqrt{\rho}\,\sigma\,\sqrt{\rho}}\,\right)^{2}, \nonumber
\end{align}

and the average taken over both the uniform time grid and the
test-trajectory ensemble. Throughout this work, \(F\) denotes the
squared Uhlmann fidelity (Jozsa convention), so that \(F=1\) for
identical states and the closed-form simplification \(F=\mathrm{Tr}(\rho\,\sigma)\)
holds when either argument is pure. When either argument is numerically pure
(purity \(> 0.999\)), we evaluate this algebraically equivalent
expression \(F = \mathrm{Tr}(\rho\,\sigma)\) for numerical stability;
otherwise the general expression above is used.

The network outputs are not constrained to be positive semidefinite, and
for predicted states that violate positivity (\(\lambda_{\min} < 0\)) the
matrix square root in Eq.~\eqref{eq:fidavg} is not directly defined. When
evaluating the general (mixed-state) fidelity we therefore clip the
negative eigenvalues of \(\rho_{\text{pred}}(t)\) to zero when forming the
factor \(\sqrt{\rho_{\text{pred}}}\), following the nearest-physical-state
prescription used in maximum-likelihood quantum state
estimation\cite{smolin2012}; the resulting fidelity is then clipped to \([0,1]\).
We separately quantify the departure of the raw network output from the
physical state space through two reported diagnostics: the most negative
eigenvalue \(\lambda_{\min}\) encountered along the trajectory, and the
total negative weight \(\sum_{i}\max(0, -\lambda_{i})\). These diagnostics
are not fed back into training except through the soft positivity penalty
of Section~3.2, so the reported \(\lambda_{\min}\) values reflect the
unconstrained network outputs.

Two diagnostic decompositions enter the analysis. First, we use a window
decomposition. We track the early-time and late-time trajectory
fidelities, defined respectively as
\(\bar{F}_{\text{early}} = \langle F\rangle_{t < T/4}\)(the
first quarter of the rollout) and
\(\bar{F}_{\text{late}} = \langle F\rangle_{t \geq 3T/4}\)(the
final quarter). The quantity
\(\Delta F = \bar{F}_{\text{late}} - \bar{F}_{\text{early}}\) is
positive when late-time predictions are more accurate than early-time
predictions, a signature consistent with the dissipative contraction of
trajectory errors that we analyze in Section 5.5.

Second, we perform completely positive trace-preserving (CPTP)
consistency checks by tracking the minimum eigenvalue
\(\lambda_{\min}\) of the predicted density operator
\(\rho_{\text{pred}}(t)\) and the trace deviation
\(\mid \mathrm{Tr}\,\rho_{\text{pred}}(t) - 1 \mid\). Neither metric enters
the training loss except through the soft positivity penalty
\(\mathcal{L}_{\text{positivity}}\) described in Section 3.2. Both are
reported as diagnostics of the departure of the predicted states from
the physical state space.

\section{Surrogate model and experimental design}

\subsection{Neural ODE surrogate architecture}

We model the open-system evolution with a Neural Ordinary Differential
Equation (NODE) surrogate. The reduced density matrix
\(\rho(t) \in \mathbb{C}^{d \times d}\) is represented as a flat vector
\(v(t) \in \mathbb{R}^{2d^{2}}\) containing the real and imaginary parts
of its entries stacked together; this representation treats the density
matrix as a general complex \(d \times d\) matrix with \(2d^{2}\) real
components. The surrogate learns a parameterized vector field
\(f_{\theta}(v)\) such that \(dv/dt = f_{\theta}(v)\); the dynamics are
autonomous, so \(f_{\theta}\) carries no explicit time dependence.

The vector \(v(0)\) is constructed from \(\rho(0)\) by stacking real and
imaginary parts, and \(\rho_{\text{pred}}(t)\) is recovered from
\(v(t)\) by the inverse map followed by Hermitization,
\(\rho \mapsto (\rho + \rho^{\dagger})/2\), so that predictions are
Hermitian by construction. Trace preservation and positivity are not
enforced architecturally; they are instead encouraged softly through the
trace and positivity penalties in the training loss (\S3.2); at
evaluation, negative eigenvalues are clipped when computing the
fidelity, as described in \S2.4.

Predictions at the test time grid are obtained by integrating
\(f_{\theta}\) forward from \(v(0)\) with the dopri5 adaptive Runge--Kutta
solver\cite{dormand1980} (relative and absolute tolerances both
\(10^{- 6}\)); training gradients are computed via the adjoint
sensitivity method\cite{chen2018node}, which avoids storing
intermediate ODE states during the backward pass.

The vector field \(f_{\theta}\) is a fully-connected feed-forward
multilayer perceptron (MLP):
four hidden layers of width 256 with sigmoid-linear-unit (SiLU)
activations and a linear
output head matching the input dimension.\cite{elfwing2018,ramachandran2017} We
treat the architecture as fixed across all
\(\left( n,\gamma,D \right)\) cells in this study. This choice shapes how
the results should be read (\S3.5, \S6).

\subsection{Training objective and the role of the trace-distance term}\label{sec:training}

All cells are trained with the composite loss

\begin{equation}
\label{eq:4}
\begin{split}
\mathcal{L}_{\mathrm{total}} = {}&\mathcal{L}_{\text{state-MSE}} + \lambda_{\mathrm{tr}}\mathcal{L}_{\text{trace-constraint}} \\
&+ \lambda_{\mathrm{pos}}\mathcal{L}_{\text{positivity}} + \lambda_{\mathrm{TD}}\mathcal{L}_{\text{trace-distance}}
\end{split}
\end{equation}

with the four terms defined as follows.
\(\mathcal{L}_{\text{state-MSE}} = \langle \| \rho_{\text{pred}}(t) - \rho_{\text{true}}(t) \|_{F}^{2}\rangle_{t,\text{batch}}\) is
the time-averaged squared Frobenius error between predicted and
reference density matrices.
\(\mathcal{L}_{\text{trace-constraint}} = \langle \lvert \mathrm{Tr}\,\rho_{\text{pred}}(t) - 1 \rvert^{2}\rangle\) is
a soft penalty enforcing unit trace, which is not guaranteed by the
reconstruction (\S3.1).
\(\mathcal{L}_{\text{positivity}} = \langle\max(0, - \lambda_{\min}(\rho_{\text{pred}}(t)))^{2}\rangle\) is
a soft hinge penalty on the most negative eigenvalue of
\(\rho_{\text{pred}}(t)\), encouraging positive-semidefinite predictions
without strictly enforcing them.
\(\mathcal{L}_{\text{trace-distance}} = \langle\frac{1}{2} \| \rho_{\text{pred}}(t) - \rho_{\text{true}}(t) \|_{1}\rangle\) is
the time-averaged quantum trace distance \(D\); because
\(\rho_{\text{pred}}(t) - \rho_{\text{true}}(t)\) is Hermitian, \(D\) is
computed as one half of the sum of the absolute eigenvalues of the
difference.

The trace distance is related to fidelity through the Fuchs--van de
Graaf inequalities, \(1 - \sqrt{F} \leq D \leq \sqrt{1 - F}\), with
\(F\) the squared fidelity of Eq.~\eqref{eq:fidavg}. Penalizing \(D\) therefore improves
the lower bound on \(F\). Unlike the Frobenius term, \(D\) remains of
order unity for the near-static predictions that collapse the
\(n = 4\) state-MSE objective (\S5.1). This persistence arises because the
trace norm sums the absolute eigenvalues of the difference and is
bounded within \(\left[ 0,1 \right]\) independently of
Hilbert-space dimension.

We fix \(\lambda_{\text{tr}} = 1.0\) and
\(\lambda_{\text{pos}} = 1.0\) across all cells. The hyperparameter
\(\lambda_{\text{TD}}\) is introduced to address the \(n = 4\) collapse
and takes two values in this work: \(\lambda_{\text{TD}} = 0\) for Phase
1 (standard training) and \(\lambda_{\text{TD}} = 0.1\) for the
fidelity-aware retraining in \S5.

Optimization uses Adam (adaptive moment estimation)\cite{kingma2014} (learning rate
\(3 \times 10^{- 4}\), default \(\beta\) coefficients) with a fixed batch
size of 32 trajectories and 300 epochs per cell. Models are evaluated on
the held-out test set every 20 epochs and the best-test-fidelity
checkpoint is retained as the trained model. All surrogate-training runs were
performed on a SLURM-managed cluster using a single NVIDIA Tesla
V100-PCIE-16GB GPU per cell, with 8 CPU cores and 48 GB of memory
allocated per job. Reference-trajectory generation (QuTiP 5
master-equation integration) runs on CPU; only Neural ODE training and
rollout use the GPU. Each cell was allotted a six-hour wall-clock cap,
with the best-test-fidelity checkpoint retained throughout training. The
largest cells, the n = 4 runs at D = 2000, reached this cap, with
elapsed times of 6.0 to 6.1 hours, stopping before the nominal 300-epoch
schedule completed, while smaller cells at lower n and D completed
within it. Because the best checkpoint is retained, cap-truncated cells
report their best achieved model rather than a partially trained final
state. Our representation-limited interpretation of the n = 4 collapse
rests on the insensitivity of fidelity to a sixteen-fold increase in
training-set size, not on epoch count, so the wall-clock cap does not
affect that conclusion.

\subsection{Controlled sweep design}\label{sec:sweep}

Phase 1 enumerates the Cartesian product of three factors. The qubit
count \(n\) takes the values \(2\), \(3\), and \(4\), fixing the
Hilbert-space dimension \(d = 2^{n}\) at \(4\), \(8\), and
\(16\) respectively. The dimensionless dissipation strength
\(\gamma/J\) ranges over \(\left\{ 0,0.01,0.1,1,10 \right\}\), spanning
four orders of magnitude together with the closed-system limit
\(\gamma = 0\). The training-set size \(D\) takes the values \(500\),
\(2000\), and \(8000\), a sixteen-fold range. The product of these three
factors yields \(45\) parameter cells; each is trained with three
independent random seeds, for \(135\) planned surrogate-training runs.
All hyperparameters not listed above are held fixed; the only controlled
variation is in the tuple \(\left( n,\gamma,D \right)\) and the
initialization seed.

In a targeted extension, we re-train the full \(n = 4\) dissipation row
at \(D = 2000\) with the fidelity-aware loss
(\(\lambda_{\text{TD}} = 0.1\)), comprising the five Phase 1 values of
\(\gamma/J\) across three seeds, and add a bounding row at
\(\gamma/J = 100\) with three further seeds to characterize the tail of
the recovery curve. This extension contributes eighteen additional
surrogates. The \(\gamma/J = 100\) point serves solely to bound the
high-dissipation tail of the \(n = 4\) recovery; it is not a complete
extension of the Phase 1 grid to \(n = 2\) or \(n = 3\), both of which
already saturate at moderate dissipation. All other hyperparameters
match Phase 1.

\subsection{Pre-committed decision rules}

To guard against post-hoc rationalization, we pre-committed to two
decision rules before executing the most resource-intensive parts of the
sweep. The first determined whether a loss-function intervention was
warranted at all: if the trajectory-averaged fidelity
\(\bar{F}_{\text{avg}}\) at the closed-system \(n = 4\) cell
exceeded \(0.6\) in two or more seeds, we would retain the Phase 1
configuration without modification and report its results as they stood.
The observed baseline, with \(\bar{F}_{\text{avg}}\) between
roughly \(0.18\) and \(0.29\) across all \(n = 4\) cells, fell short of
this threshold and triggered the fidelity-aware loss intervention
detailed in \S5. The second rule governed whether to commit resources to
the full fidelity-aware row. Following a three-seed pilot at \(n = 4\),
\(\gamma/J = 0.1\), and \(\lambda_{\text{TD}} = 0.1\), the extension was
conditioned on \(\bar{F}_{\text{avg}} \geq 0.55\) in at least
two of three seeds, with the late-time advantage
\(\bar{F}_{\text{late}} > \bar{F}_{\text{early}}\) restored.
The pilot produced
\(\bar{F}_{\text{avg}} = 0.601 \pm 0.024\) with the late-time
advantage present in all three seeds, satisfying the condition and
triggering the full row. The full row, trained as an independent set of
runs, reproduced this recovery at
\(\bar{F}_{\text{avg}} = 0.608 \pm 0.003\) for the same cell
(Table~\ref{tab:recovery}).

We emphasize this pre-commitment because the \(n = 4\) result was the
central investigative question of the study. A successful loss
modification could otherwise be interpreted as tuning the objective
after observing the baseline outcome. Fixing both the numerical
threshold and the structural decision (the full row,
\(\lambda_{\text{TD}} = 0.1\), three seeds) before executing each stage
helps mitigate that risk.

\subsection{What the design does and does not control for}

The sweep controls three sources of variation. Training-set size is
varied over a sixteen-fold range through \(D \in \{ 500,2000,8000\}\).
Seed variability is quantified using three independent seeds per cell
and the corresponding standard deviations. The loss objective is
isolated by ensuring that the Phase 1 sweep and the fidelity-aware
extension differ only in \(\lambda_{\text{TD}}\), with all other
training settings held fixed.

Several factors are deliberately held constant, so that our conclusions
remain conditioned upon them. We test a single NODE architecture, a
single integration solver (dopri5), a single optimizer (Adam), a fixed
trajectory horizon (\(T = 10/J\)), and a single training-time
initial-state distribution (Haar-random pure states, evaluated against
the broader test ensemble of \S2.3).

\section{Results: n = 2 and n = 3}

\begin{figure}[t]
\centering
\includegraphics[width=\columnwidth]{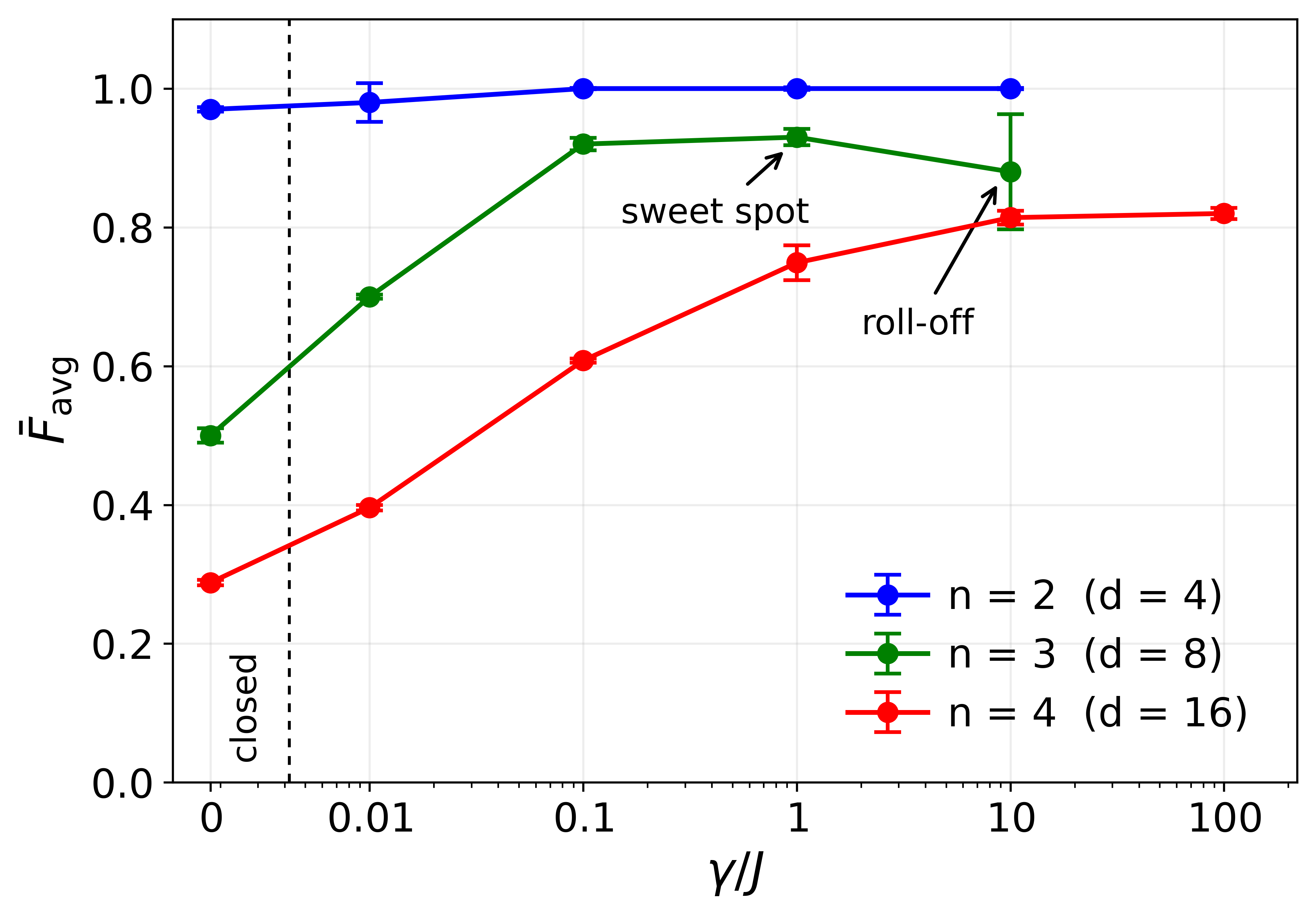}
\caption{Trajectory-averaged fidelity $\bar{F}_{\mathrm{avg}}$ versus
dimensionless dissipation strength $\gamma/J$ at $D = 2000$, one line per
system size $n$ (markers show three-seed means; error bars show the seed
standard deviation). The $n=2$ and $n=3$ curves use standard state-MSE
training; the $n=4$ curve uses the fidelity-aware loss
($\lambda_{\mathrm{TD}} = 0.1$; \S\ref{sec:training}) and includes the
$\gamma/J = 100$ bounding point. $n=2$ saturates near unity; $n=3$ shows
a broad sweet spot at $\gamma/J \in \{0.1,1\}$ with a slight roll-off at
$\gamma/J = 10$; $n=4$ recovers monotonically and saturates by
$\gamma/J = 100$, with the favorable regime shifted toward stronger
damping relative to $n=3$. The closed-system ($\gamma = 0$) point is
offset to the left of the logarithmic axis and separated by a dashed
divider.}
\label{fig:favg_overview}
\end{figure}

Figure~\ref{fig:favg_overview} provides an overview of the central
result: trajectory-averaged fidelity $\bar{F}_{\mathrm{avg}}$ as a
function of dissipation strength $\gamma/J$ for each system size. The
trends summarized there---near-saturation at $n=2$, an interior sweet
spot at $n=3$, and a damping-shifted recovery at $n=4$---are developed
in detail in this section and \S\ref{sec:n4}.

\begin{figure*}[t]
\centering
\includegraphics[width=\textwidth]{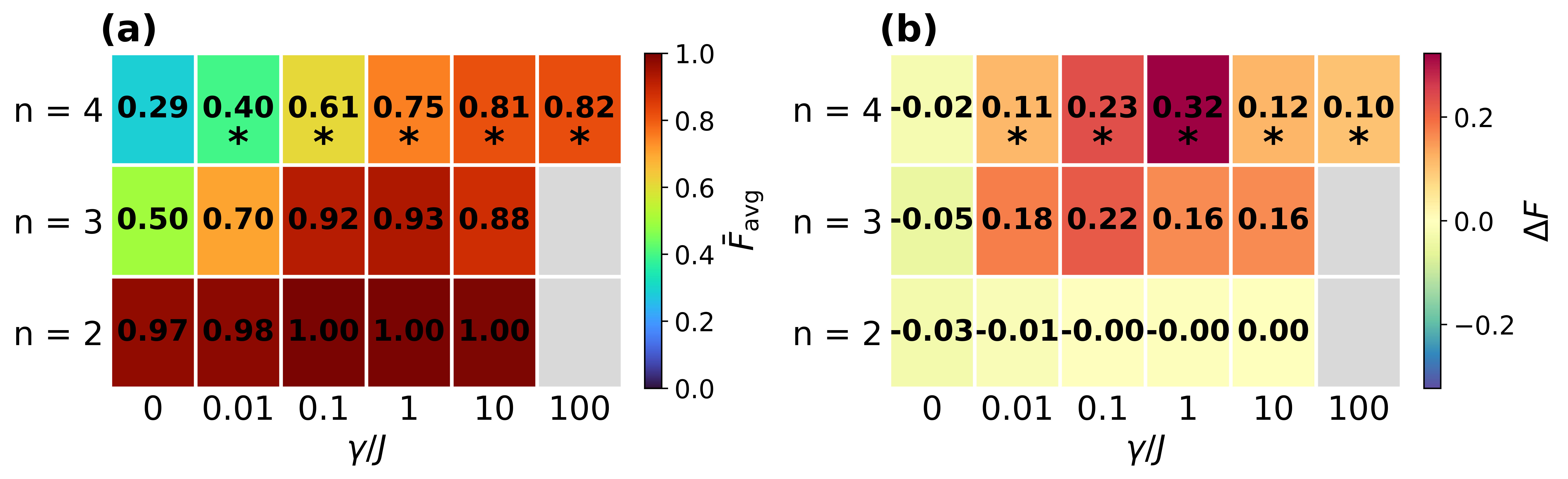}
\caption{Neural ODE learnability across the $n \times \gamma$ sweep.
\textbf{(a)}~Trajectory-averaged fidelity $\bar{F}_{\mathrm{avg}}$.
\textbf{(b)}~The late-versus-early fidelity gap
$\Delta F = \bar{F}_{\mathrm{late}} - \bar{F}_{\mathrm{early}}$;
positive values indicate higher accuracy in the late-time window.
Cells marked with an asterisk ($*$) at $n=4$ were trained with the
fidelity-aware loss ($\lambda_{\mathrm{TD}} = 0.1$; \S\ref{sec:training}); all other
cells use standard state-MSE training. The $\gamma/J = 100$ column is a
separately added bounding experiment for the $n=4$ row only (\S\ref{sec:sweep}); it
is left blank at $n=2$ and $n=3$. Blank (grey) cells indicate combinations that
were not simulated. See \S\ref{sec:sweep} for cell counts.}
\label{fig:learnability_sweep}
\end{figure*}

\subsection{n = 2 is essentially saturated}

At \(n = 2\)(\(d = 4\)) the NODE surrogate is essentially saturated. At
\(D = 2000\) it achieves
\(\bar{F}_{\text{avg}} \geq 0.97\) across the entire
\(\gamma/J\) range, with the closed-system cell (\(\gamma = 0\)) at
\(\bar{F}_{\text{avg}} = 0.97\) and the most strongly
dissipative cell (\(\gamma/J = 10\)) reaching
\(\bar{F}_{\text{avg}} = 1.00\) within seed variability. At the
smallest budget \(D = 500\) a few cells dip lower, to trajectory-averaged
fidelities between \(0.89\) and \(0.94\), but fidelity remains high
throughout.

The window decomposition shows
\(\Delta F = \bar{F}_{\text{late}} - \bar{F}_{\text{early}}\) statistically
indistinguishable from zero at every dissipation strength
(\(\mid \Delta F \mid \leq 0.03\)), so early-time and late-time
predictions are comparably accurate. We observe high fidelity with
little dependence on \(D\), with no systematic improvement from
\(D = 500\) to \(D = 8000\). The \(n = 2\) system serves primarily as a
control regime, confirming that the architecture, training pipeline, and
evaluation harness achieve near-saturated performance when the learning
problem is comparatively easy.

\subsection{n = 3 reveals dissipation-shaped learnability}

At \(n = 3\)(\(d = 8\)) the first substantial degradation appears:
\(\bar{F}_{\text{avg}}\) drops to \(\approx 0.50\) in the
closed-system cell, climbs to a plateau of \(\approx 0.92\) at
\(\gamma/J \in \{ 0.1,1.0\}\), and falls slightly to \(\approx 0.88\) at
\(\gamma/J = 10\). Unlike the static-prediction collapse observed at
\(n = 4\)(\S5.1), the closed-system \(n = 3\) surrogate continues to
reduce its training loss past epoch 20 and produces non-trivial temporal
variation in its predictions. The fidelity shortfall is therefore
consistent with accumulated divergence between predicted and reference
trajectories over the integration horizon, rather than a failure to
learn any dynamics at all.

The dissipative cells exhibit a pronounced late-time accuracy advantage:
\(\Delta F \approx + 0.18\) at \(\gamma/J = 0.01\) and \(+ 0.16\) to
\(+ 0.22\) across \(\gamma/J \in \{ 0.1,1,10\}\), with
\(\bar{F}_{\text{late}} > \bar{F}_{\text{early}}\) in
all three seeds at every dissipative configuration. We interpret this as
evidence consistent with dissipative contraction of rollout errors: the
surrogate is most accurate in the late-time portion of the trajectory,
where the open-system dynamics suppress errors accumulated earlier in
the rollout. This is the late-time accuracy signature that the
\(n = 4\) fidelity-aware-loss intervention subsequently restores (\S5).

The dissipation-assisted improvement is not attributable to training-set
size. Across \(D \in \{ 500,2000,8000\}\) at \(n = 3\),
\(\bar{F}_{\text{avg}}\) varies by at most \(\approx 0.06\) at
fixed \(\gamma\): a sixteen-fold increase in training data does not
close the gap between the closed and dissipative cells, whereas
introducing weak dissipation (\(\gamma/J = 0.1\)) at the smallest budget
\(D = 500\) already lifts \(\bar{F}_{\text{avg}}\) to
\(\approx 0.86\), rising only to \(\approx 0.92\) at \(D = 2000\). This
weak dependence on \(D\) suggests the closed-system shortfall is not
primarily data-limited.

A slight roll-off appears at the strongest dissipation. At
\(\gamma/J = 10\)(\(n = 3\)),
\(\bar{F}_{\text{avg}} \approx 0.88\) sits below the
\(\approx 0.92\) plateau of \(\gamma/J \in \{ 0.1,1\}\), though with
substantial seed spread. We interpret this as the onset of an
over-damped regime in which trajectories rapidly approach the steady
state. The single-row evidence is too sparse to separate this from seed
noise, but the \(n = 3\) results remain consistent with an interior
performance sweet-spot in \(\gamma\). At \(n = 4\)(\S5) the
highest-fidelity regime instead shifts toward larger \(\gamma\) without
an interior peak, a contrast we develop there.

\section{Results: n = 4 under fidelity-aware training}\label{sec:n4}

\subsection{n = 4 closed-system dynamics resist MSE-trained surrogates}

Under standard state-MSE training, Neural ODE surrogates at
\(n = 4\) converge to a near-static prediction whose state-MSE is low but
whose fidelity is poor. Across all 15 Phase 1 cells at \(n = 4\)(45
runs: \(\gamma/J \in \{ 0,0.01,0.1,1,10\}\),
\(D \in \{ 500,2000,8000\}\), three seeds), trajectory-averaged fidelity
remains in the \(0.18\) to \(0.29\) range, below the closed-system
\(n = 3\) baseline of \(\approx 0.50\), and largely insensitive to a
sixteen-fold increase in training-set size
(Figure~S1 in the Supplementary Material). Three independent
diagnostics suggest that this failure is not primarily attributable to
data scarcity.

\begin{figure*}[t]
\centering
\includegraphics[width=\textwidth]{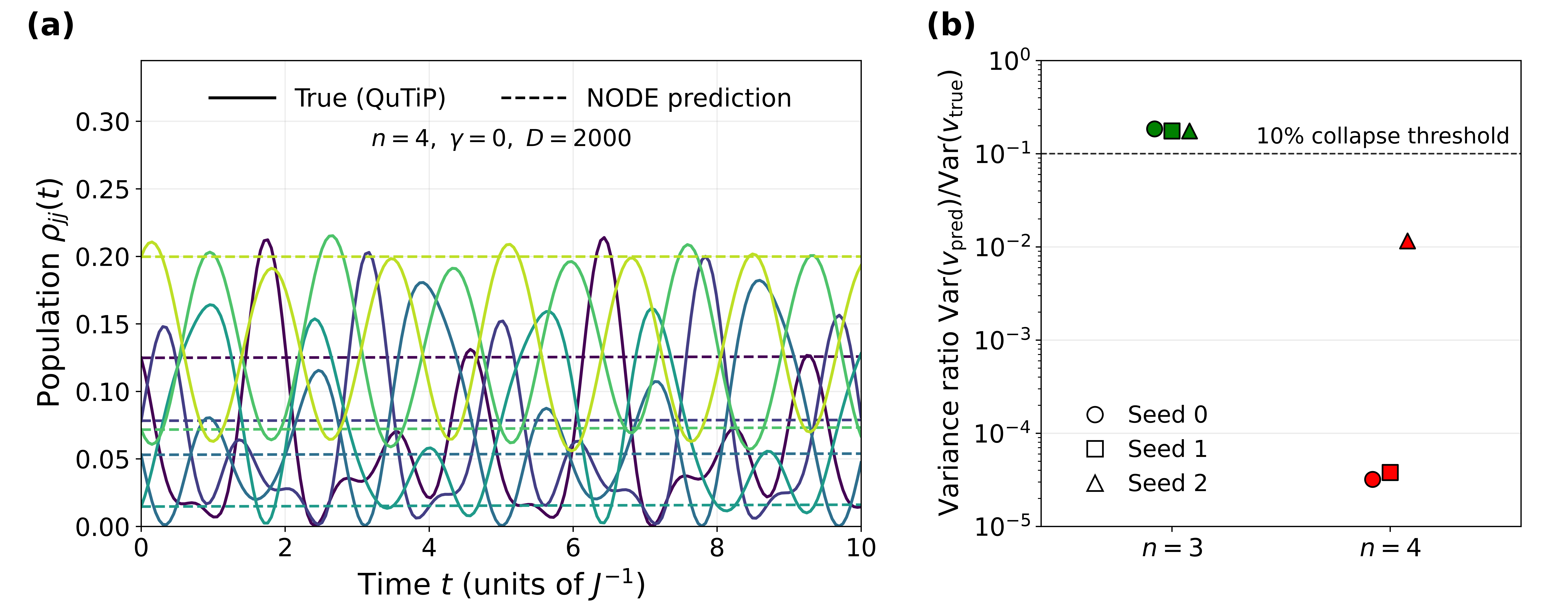}
\caption{Closed-system ($\gamma = 0$) static-prediction collapse at
$n = 4$. \textbf{(a)}~True (solid) versus predicted (dashed)
density-matrix populations $\rho_{jj}(t)$ for a representative $n = 4$
test trajectory (the six most dynamically varying populations are
shown); the true populations oscillate while the prediction is
essentially frozen. \textbf{(b)}~Flat-vector temporal-variance ratio
$\mathrm{Var}(v_{\mathrm{pred}})/\mathrm{Var}(v_{\mathrm{true}})$ across
three seeds for $n = 3$ (partial learning, $\approx 18\%$) and $n = 4$
(static collapse, $0.003$--$1.16\%$); bars show seed means and points
show individual seeds. The dashed line marks the $10\%$ collapse
threshold of the diagnostic.}
\label{fig:static_collapse}
\end{figure*}

First, the training loss rapidly plateaus: all \(n = 4\) cells reach
\(\mathcal{L}_{\text{total}}\) between \(0.002\) and \(0.004\) by epoch 20
and remain near that level thereafter, converging to a low-loss basin
associated with poor fidelity. Second, the predictions exhibit a static
signature: using the temporal variance of the predicted flat-vector
state representation, averaged over the test ensemble, the
variance-retention ratio
\(\mathrm{Var}(v_{\text{pred}})/\mathrm{Var}(v_{\text{true}})\) is
approximately \(18\%\) at \(n = 3\) across three seeds but ranges from
\(0.003\%\) to \(1.2\%\) at \(n = 4\), so the \(n = 4\) model output is
nearly time-invariant in every seed (Figure~\ref{fig:static_collapse}). Third, the
dissipative-contraction signature is absent: whereas
\(n = 3\) dissipative cells satisfy
\(\bar{F}_{\text{late}} > \bar{F}_{\text{early}}\),
all \(n = 4\) cells satisfy
\(\bar{F}_{\text{late}} \leq \bar{F}_{\text{early}}\).

We attribute this behavior to the divergence between state-MSE and
quantum fidelity that motivated the trace-distance term (\S3.2), which
becomes particularly consequential at \(d = 16\). There, the trivial
static prediction \(\rho_{\text{pred}}(t) \equiv \rho(0)\) achieves
state-MSE \(\approx 6 \times 10^{- 4}\) against trajectories that evolve
toward the steady state, comparable to values obtained by well-trained
surrogates at \(n = 2\) and \(n = 3\). The state-MSE objective therefore
admits a low-loss basin of near-static predictions at large \(d\), from
which the optimizer does not escape. A targeted diagnostic confirms this
failure mode is specific to \(n = 4\): at \(n = 3\), \(\gamma = 0\), the
training loss continues to decrease substantially beyond epoch 20 in
every seed (from \(\mathcal{L}_{\text{total}} \approx 0.0104\) at epoch
20 to \(\approx 0.0046\) by epoch 100, a 55--56\% further descent in all
three seeds), consistent with partial learning accompanied by
rollout-error accumulation rather than the static collapse seen at
\(n = 4\).

\subsection{Fidelity-aware training escapes the static minimum}

Having identified the static-collapse pathology, we ask whether it
reflects a fundamental limitation of the dynamics or a consequence of
the training objective. We therefore treat \(\lambda_{\text{TD}}\) as a
hyperparameter and select its value through a pilot study. We first
screened \(\lambda_{\text{TD}} = 0.01\) on the \(n = 4\),
\(\gamma/J = 0.1\), \(D = 2000\) cell with a single seed. The model
failed to escape the static-prediction basin, yielding
\(\bar{F}_{\text{avg}} = 0.232\) with
\(\bar{F}_{\text{late}} = 0.231\) below
\(\bar{F}_{\text{early}} = 0.294\), no late-time advantage and
a fidelity below the MSE-only \(n = 4\) baseline of
\(\approx 0.29\)(\S5.1). Because this single-seed screen failed
decisively, we did not expand it to three seeds, and instead advanced
the next candidate value to a full pilot. At
\(\lambda_{\text{TD}} = 0.1\), evaluated across three seeds, the model
recovered learnability:
\(\bar{F}_{\text{avg}} = 0.601 \pm 0.024\), with
\(\bar{F}_{\text{late}} > \bar{F}_{\text{early}}\) restored
in all three seeds.

The improvement is substantial: increasing \(\lambda_{\text{TD}}\) from
\(0.01\) to \(0.1\) moves the model from a regime indistinguishable from
MSE-only training to one in which the late-time advantage is restored.
This comes at a measurable cost in positivity, however. The minimum
eigenvalue deteriorates from \(\approx - 0.019\) at
\(\lambda_{\text{TD}} = 0.01\) to \(\approx - 0.105\) at
\(\lambda_{\text{TD}} = 0.1\), a trade-off discussed in \S5.4.

We adopt \(\lambda_{\text{TD}} = 0.1\) for the full \(n = 4\gamma\)-row
reported below. No further tuning of \(\lambda_{\text{TD}}\) was
performed. Because we did not conduct a complete
\(\lambda_{\text{TD}}\) sweep, this value should be regarded as a
diagnostic choice that substantially improves fidelity rather than an
optimized hyperparameter.

\subsection{Recovery across the \(\gamma\)-row}

Re-training the full \(n = 4\gamma\)-row at \(D = 2000\) with
\(\lambda_{\text{TD}} = 0.1\)(15 cells: 5 \(\gamma\) values \(\times\) 3 seeds;
see \S3) recovers learnability across the entire dissipative regime
tested. Table~\ref{tab:recovery} reports per-\(\gamma\) trajectory-averaged fidelity,

\begin{table*}
\centering
\caption{Neural ODE recovery across the $n=4$ dissipation row under fidelity-aware training ($D=2000$, $\lambda_{\mathrm{TD}}=0.1$, three seeds per $\gamma$). $\bar{F}$ values are mean $\pm$ sample standard deviation across seeds. The baseline columns report the higher of two trivial-predictor fidelities evaluated on the same 100-trajectory mixed test ensemble (\S\ref{sec:initial}): the static predictor $\rho_{\mathrm{pred}}(t)\equiv\rho(0)$ and the steady-state predictor $\rho_{\mathrm{pred}}(t)\equiv|0\dots0\rangle\langle0\dots0|$; the Type column indicates which was higher at that $\gamma$. $\lambda_{\min}$ is the seed-averaged most-negative eigenvalue of $\rho_{\mathrm{pred}}$ encountered along the trajectory (computed on the raw network output, before eigenvalue clipping). Because $\bar{F}_{\mathrm{avg}}$ averages over the full trajectory whereas $\bar{F}_{\mathrm{early}}$ and $\bar{F}_{\mathrm{late}}$ are the first- and final-quarter windows (\S\ref{sec:fidelity}), $\bar{F}_{\mathrm{avg}}$ need not lie between them, as in the $\gamma=0$ row. The $\gamma/J=100$ row is a separately added bounding experiment (\S\ref{sec:sweep}), not part of the 15-cell Phase~1 row.}
\label{tab:recovery}
\begin{tabular}{lcccccc}
\hline\hline
$\gamma/J$ & $\bar{F}_{\mathrm{avg}}$ & $\bar{F}_{\mathrm{early}}$ & $\bar{F}_{\mathrm{late}}$ & Late\,$>$\,Early & Baseline $\bar{F}$ (Type) & $\lambda_{\min}$ \\
\hline
$0$    & $0.288 \pm 0.004$ & $0.348$ & $0.328$ & $0/3$ & $0.288$ (Static $\rho(0)$)   & $-0.008$ \\
$0.01$ & $0.396 \pm 0.004$ & $0.377$ & $0.492$ & $3/3$ & $0.238$ (Static $\rho(0)$)   & $-0.036$ \\
$0.1$  & $0.608 \pm 0.003$ & $0.489$ & $0.722$ & $3/3$ & $0.258$ (Steady-state)       & $-0.102$ \\
$1$    & $0.749 \pm 0.025$ & $0.540$ & $0.863$ & $3/3$ & $0.860$ (Steady-state)       & $-0.102$ \\
$10$   & $0.814 \pm 0.010$ & $0.733$ & $0.850$ & $3/3$ & $0.984$ (Steady-state)       & $-0.100$ \\
$100$  & $0.820 \pm 0.008$ & $0.750$ & $0.852$ & $3/3$ & $0.995$ (Steady-state)       & $-0.102$ \\
\hline\hline
\end{tabular}
\end{table*}

early- and late-window means, and the unconstrained minimum eigenvalue
\(\lambda_{\min}\) of the predicted density operator; it also includes
the separately added bounding row at \(\gamma/J = 100\)(\S3.3).

Three features are notable. First, the closed-system cell remains poorly
learned (\(\bar{F}_{\text{avg}} = 0.29\), no late-time
advantage), consistent with the fidelity-aware loss alone being
insufficient to rescue learning in the absence of dissipative
contraction. Second, dissipative cells satisfy
\(\bar{F}_{\text{late}} > \bar{F}_{\text{early}}\) in
all three seeds at every \(\gamma\), restoring the late-time advantage
seen at \(n = 3\). Third, \(\bar{F}_{\text{avg}}\) increases
monotonically across \(\gamma/J \in \{ 0.01,0.1,1,10\}\)(0.39, 0.61,
0.75, 0.81) and then saturates by \(\gamma/J = 100\), where the bounding
experiment gives \(\bar{F}_{\text{avg}} = 0.820 \pm 0.008\),
indistinguishable from \(\gamma/J = 10\) within seed variability. The
observed saturation suggests that further increases in dissipation alone
are insufficient to improve performance under the present architecture
and training protocol. This contrasts with \(n = 3\), where
\(\gamma/J = 10\) already shows a slight roll-off below the
\(\gamma/J = 1\) value; at \(n = 4\) the highest-fidelity regime shifts to
higher \(\gamma\) and is bounded by a plateau rather than an interior
peak.

The window metric
\(\Delta F = \bar{F}_{\text{late}} - \bar{F}_{\text{early}}\) is
non-monotonic, peaking at \(\gamma/J = 1\)(\(\Delta F = + 0.32\)) and
falling to \(+ 0.12\) at \(\gamma/J = 10\) and \(+ 0.10\) at
\(\gamma/J = 100\), even as \(\bar{F}_{\text{avg}}\) continues
to rise. The rescue therefore separates into two regimes: at
\(\gamma/J \lesssim 1\) the gain is concentrated in the late-time window,
consistent with dissipative contraction of rollout error; at
\(\gamma/J \gtrsim 10\) the dynamics are overdamped, accuracy is roughly
uniform across the trajectory (small \(\Delta F\)), and the higher
\(\bar{F}_{\text{avg}}\) reflects a simplified trajectory rather
than richer learned dynamics.

\begin{figure}[t]
\centering
\includegraphics[width=\columnwidth]{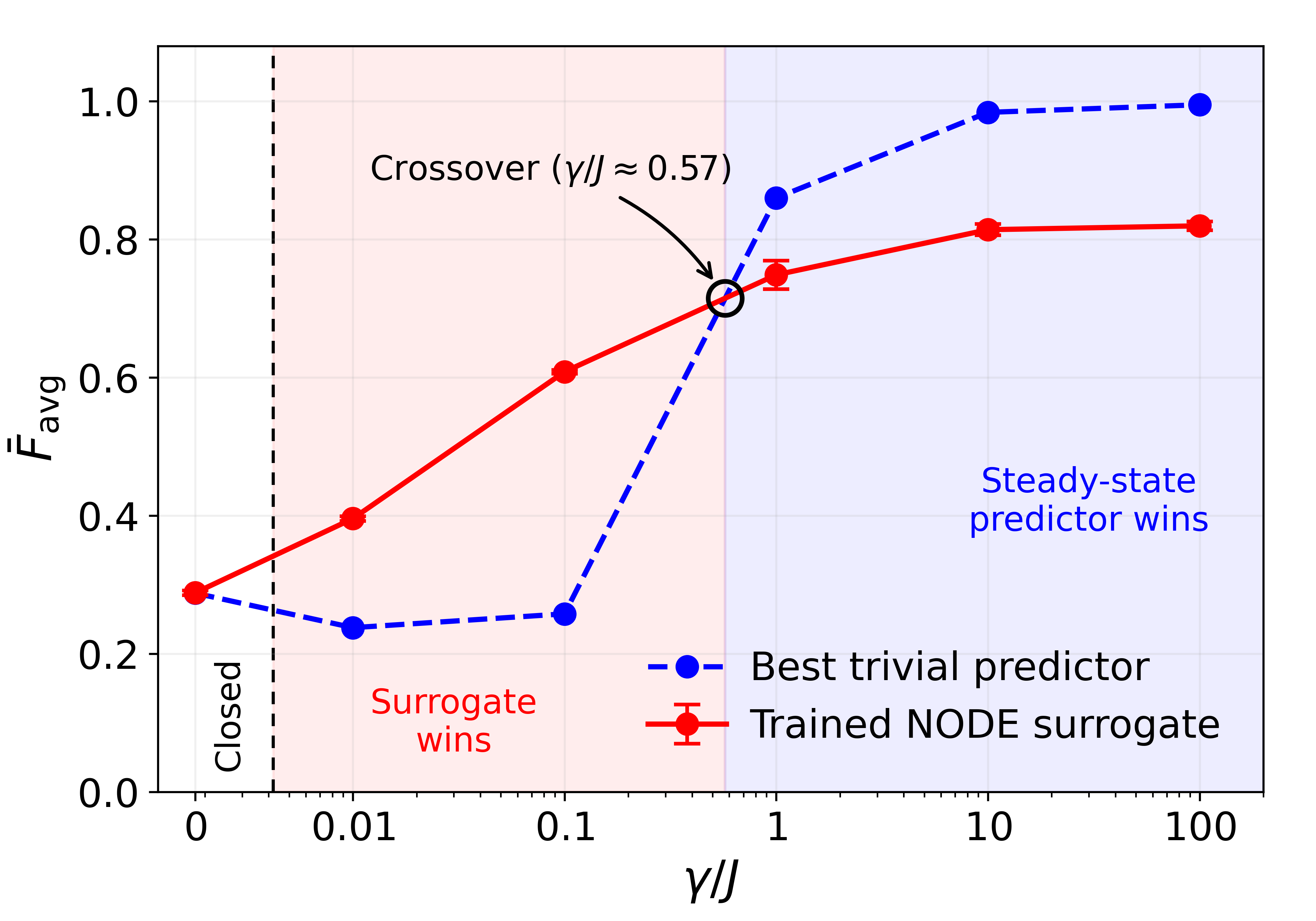}
\caption{Trained NODE surrogate $\bar{F}_{\mathrm{avg}}$ (solid, with
seed standard deviation) versus the best trivial predictor at each
$\gamma$ (dashed; the higher of the static $\rho(0)$ and steady-state
$|0\dots0\rangle\langle0\dots0|$ baselines, per Table~\ref{tab:recovery}),
for $n = 4$ at $D = 2000$. The curves cross near $\gamma/J \approx 0.57$
(interpolated in $\log\gamma$ between the $\gamma/J = 0.1$ and
$\gamma/J = 1$ points):
the surrogate adds value over trivial predictors at weak-to-moderate
dissipation (left), and is matched or overtaken by the steady-state
predictor under strong dissipation (right), where high
$\bar{F}_{\mathrm{avg}}$ reflects trajectory simplification rather than
richer learned dynamics.}
\label{fig:crossover}
\end{figure}

The trivial-predictor baselines make this distinction concrete
(Figure~\ref{fig:crossover}). At
\(\gamma = 0\), the trained
\(\bar{F}_{\text{avg}} = 0.288\) matches the static-predictor
baseline to three significant figures (both 0.288), a direct
confirmation of the static collapse of \S5.1: the surrogate has converged
to the trivial prediction \(\rho(0)\). At weak-to-moderate dissipation
(\(\gamma/J \in \{ 0.01,0.1\}\)) the trained model substantially exceeds
both baselines (by \(+ 0.16\) and \(+ 0.35\)), the regime where it learns
transient dynamics no constant predictor captures. At strong dissipation
(\(\gamma/J \geq 1\)) the comparison reverses: the steady-state
predictor outperforms the trained model by \(0.11\) to \(0.18\), reaching
\(\bar{F} = 0.995\) at \(\gamma/J = 100\). Here the trajectory is
dominated by rapid contraction to the all-zero steady state, so the
time-averaged objective is dominated by the long steady-state tail,
reducing the marginal value of learning transient structure: the trained
model achieves what a zero-parameter steady-state predictor already
provides.

The trained \(\bar{F}_{\text{avg}} \approx 0.82\) in this regime
therefore reflects physical trajectory simplification more than the
surrogate having learned richer dynamics than the trivial guess. The
late-time advantage \(\Delta F\)(peaking at \(\gamma/J = 1\)) thus
identifies the regime in which the surrogate provides predictive value
beyond trivial baselines, while the saturation of
\(\bar{F}_{\text{avg}}\) at \(\gamma/J \geq 10\) reflects
steady-state domination of the trajectory rather than improvement in
learned dynamics.

\subsection{Caveats and the positivity trade-off}

\begin{figure}[t]
\centering
\includegraphics[width=\columnwidth]{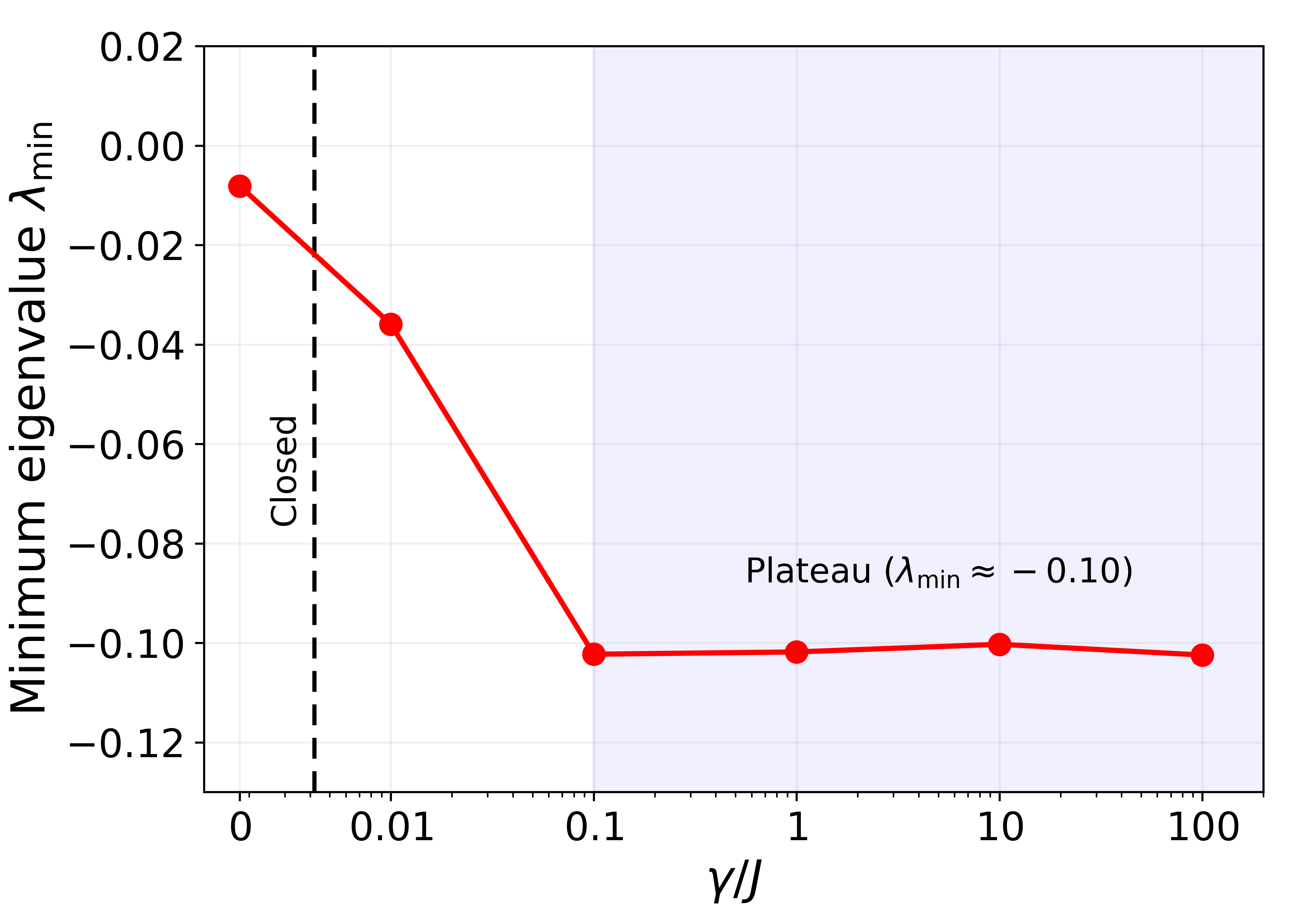}
\caption{Most-negative predicted eigenvalue $\lambda_{\min}$
(seed-averaged) along the trajectory across the $n = 4$ fidelity-aware
dissipation row. The violation deepens from $\gamma = 0$ to
$\gamma/J = 0.1$, then plateaus near $-0.10$, changing by only $0.002$
between $\gamma/J = 10$ and $\gamma/J = 100$ despite a tenfold further
increase in dissipation---consistent with a fixed architectural cost
rather than a dissipation-rate-controlled artifact. The closed-system
($\gamma = 0$) point is offset to the left of the logarithmic axis.}
\label{fig:lmin}
\end{figure}

Three caveats accompany the \(n = 4\) result. First, the fidelity-aware
loss recovers fidelity at the cost of predicted-state positivity: the
most negative eigenvalue of \(\rho_{\text{pred}}\) along the trajectory
drops from \(\approx - 0.008\) at \(\gamma = 0\), through
\(\approx - 0.036\) at \(\gamma/J = 0.01\), to \(\approx - 0.10\) for
\(\gamma/J \geq 0.1\)(Table~\ref{tab:recovery}), so the predicted states leave the
physical state space rather than remaining marginally non-physical. A
structural observation sharpens this: the violation flatlines between
\(\gamma/J = 10\)(\(\lambda_{\min} = - 0.100\)) and
\(\gamma/J = 100\)(\(\lambda_{\min} = - 0.102\)) despite a tenfold
further increase in dissipation (Figure~\ref{fig:lmin}). The observed plateau suggests that the
non-physical artifact is not primarily controlled by the dissipation
rate; it is consistent with a fixed cost incurred by the unconstrained
MLP vector field when fitting the \(d = 16\) trajectories in regimes
where dissipative contraction is substantial. We treat CPTP-constrained
surrogate architectures (such as Gram--Hadamard density operators or
purification-based representations; \S6) as the natural structural fix
and a necessary direction for future work.

Second, the rescue is established with a single class of architecture (a
Neural ODE with a feed-forward vector field); whether the same
trace-distance intervention is needed, or effective, for other surrogate
classes, including Transformer-based architectures, remains untested.
Our objective is not to benchmark architectures but to use a fixed
surrogate class as a controlled probe of how dissipation reshapes
learnability.

Third, the \(n = 3\) versus \(n = 4\) contrast (the shift of the favorable
\(\bar{F}_{\text{avg}}\) plateau toward stronger damping) is
based on only two system sizes; we do not claim a scaling law for the
optimal \(\gamma\) with Hilbert-space dimension. Moreover, while the
trained model reaches \(\bar{F}_{\text{avg}} \approx 0.82\) at
\(\gamma/J \geq 10\), the baseline comparison (Table~\ref{tab:recovery}) shows a
zero-parameter steady-state predictor reaching
\(\bar{F} = 0.98\)--\(0.99\) in the same regime; the high trained
fidelity at strong dissipation should therefore be read not as the model
having learned rich dynamics, but as the trajectory itself having
simplified to a regime where trivial predictors do well.

\subsection{Putting the rows together}

Figure~\ref{fig:learnability_sweep} summarizes the three \(n \times \gamma\) rows as a heatmap of
$\bar{F}_{\mathrm{avg}}$ (left) and \(\Delta F\)(right). Three
structural observations emerge. At \(n = 2\) the surrogate is essentially
saturated at fidelity \(\approx 1.00\) across the entire \(\gamma\) range,
so dissipation is largely irrelevant to learnability at this scale. At
\(n = 3\), closed dynamics are partially learned but rollout error
accumulates (\(\bar{F}_{\text{avg}} \approx 0.50\), no
late-time advantage); dissipation lifts the row to a broad high-fidelity
plateau at \(\gamma/J \in \{ 0.1,1.0\}\), with the late-time advantage
appearing as \(\Delta F \approx + 0.16\) to \(+ 0.22\) across
\(\gamma/J \geq 0.01\). At \(n = 4\), closed dynamics fail under
state-MSE training via static-prediction collapse, a qualitatively
distinct failure from \(n = 3\)(\S5.1); fidelity-aware training recovers
learnability monotonically across the dissipative range, with the
late-time advantage concentrated at \(\gamma/J = 1\).

Taken together, these patterns support our central claim: for the tested
Neural ODE surrogate, dissipation alters the learnability of open
quantum dynamics in a way that depends on system size. It is largely
irrelevant at \(n = 2\), gives rise to a broad high-fidelity regime at
\(n = 3\), and at \(n = 4\) enables recovery of learnable rollout
structure that is absent under closed-system training. The recovery is
most informative in the weak-to-moderate dissipation regime
(\(\gamma/J \in \{0.01, 0.1\}\)), where the trained model exceeds trivial
baselines by a clear margin; the late-time advantage itself peaks at
\(\gamma/J = 1\), though there the steady-state baseline already matches
or exceeds the trained model. At stronger dissipation the trained
fidelity is matched or exceeded by the zero-parameter steady-state
predictor, so the high fidelity there reflects trajectory simplification
rather than richer learned dynamics.

\section{Discussion}

\subsection{What these results establish}

Across the controlled sweep, five results are established for the tested
Neural ODE surrogate. First, closed-system learnability degrades with
system size. Second, dissipation systematically alters learnability.
Third, the favorable dissipation regime shifts toward stronger damping
as system size increases from \(n = 3\) to \(n = 4\). Fourth, a state-MSE
objective induces a static-prediction collapse at \(n = 4\). Fifth, a
fidelity-aware loss restores learnability and recovers the late-time
advantage observed in the dissipative \(n = 3\) regime.

\begin{figure}[t]
\centering
\includegraphics[width=\columnwidth]{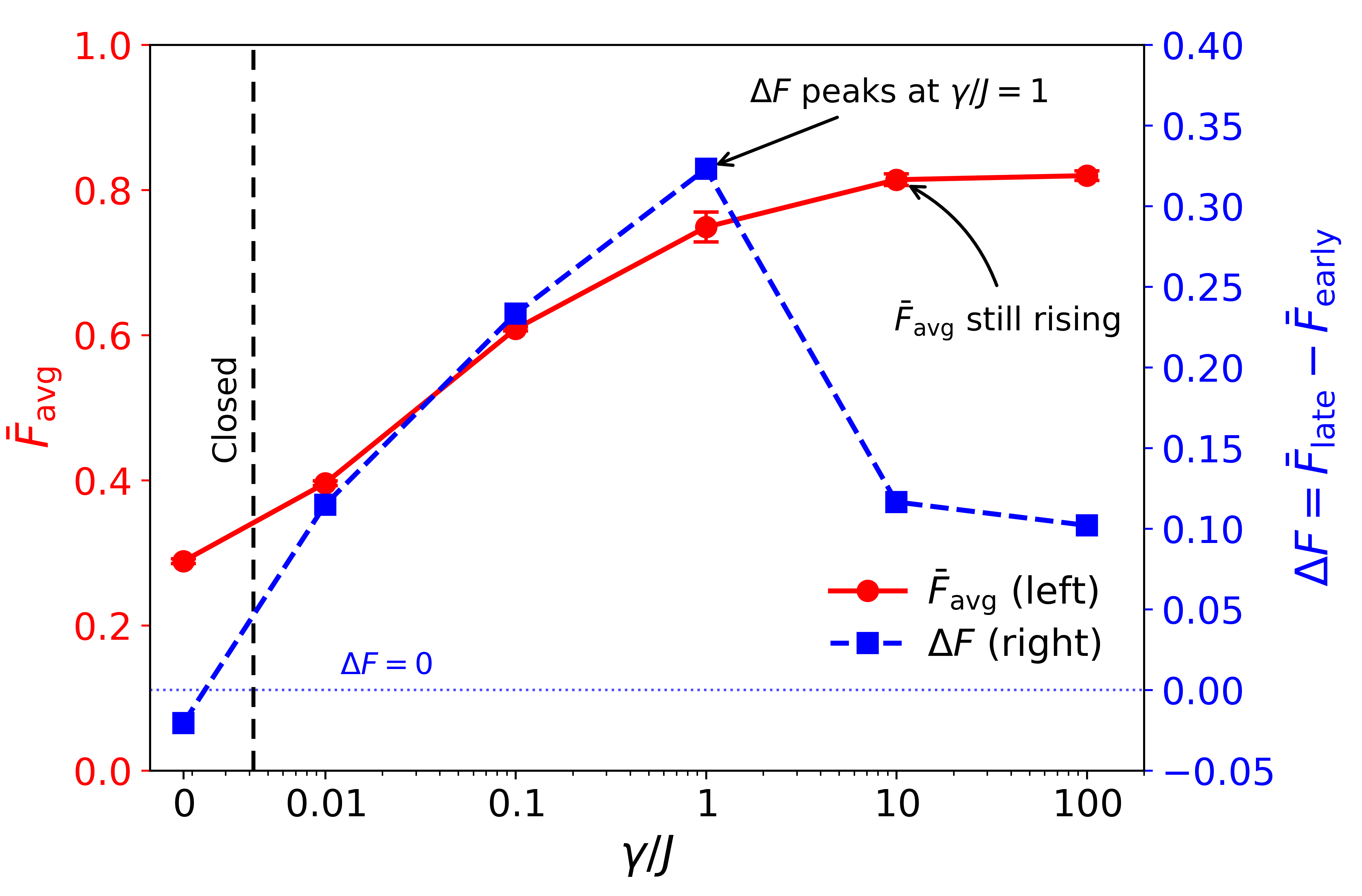}
\caption{Trajectory-averaged fidelity $\bar{F}_{\mathrm{avg}}$ (left
axis, red) and the late-versus-early gap
$\Delta F = \bar{F}_{\mathrm{late}} - \bar{F}_{\mathrm{early}}$ (right
axis, blue) across the $n = 4$ fidelity-aware dissipation row
($D = 2000$, $\lambda_{\mathrm{TD}} = 0.1$, three seeds). $\Delta F$
peaks at $\gamma/J = 1$ and then declines, while
$\bar{F}_{\mathrm{avg}}$ continues to rise and saturate---direct
evidence that dissipative contraction (peak $\Delta F$) and trajectory
simplification (rising $\bar{F}_{\mathrm{avg}}$) are distinct effects
that peak at different dissipation strengths.}
\label{fig:overlay}
\end{figure}

These observations show that, for the tested surrogate, the effect of
dissipation on learnability depends on both system size and dissipation
strength. The recovery at \(n = 4\) is most informative at
weak-to-moderate dissipation (\(\gamma/J \in \{0.01, 0.1\}\)), where the
trained model substantially exceeds both static and steady-state
baselines. At stronger damping (\(\gamma/J \geq 1\)) the dynamics become
increasingly dominated by contraction toward the steady state, and the
zero-parameter steady-state predictor matches or exceeds the trained
model. The late-time advantage \(\Delta F\) peaks near
\(\gamma/J = 1\) while trajectory-averaged fidelity saturates only at
larger \(\gamma\), indicating that dissipative contraction and
trajectory simplification are distinct effects that do not coincide
(Figure~\ref{fig:overlay}).

The same pattern appears wherever the late-time advantage is observed:
dissipation produces a positive
\(\Delta F = \bar{F}_{\text{late}} - \bar{F}_{\text{early}}\),
consistent with the surrogate capturing contraction toward the steady
state. It is absent in the \(n = 2\) row, where the trajectory is learned
uniformly well in time, and at \(n = 4\), \(\gamma = 0\) even under
fidelity-aware training, where no contraction exists to exploit.

\subsection{Limitations}

Several limitations bound these conclusions. First, all results derive
from a single Neural ODE configuration (MLP vector field, four hidden
layers of width 256, SiLU activations, dopri5 integration); an inductive
bias better suited to closed coherent dynamics (Hamiltonian-structured
Neural ODEs, symplectic networks, or Transformer-based trajectory
models) may behave differently in the closed-system regime, and we
regard such architecture-independence checks as an important follow-up.

Second, fidelity recovery at \(n = 4\) comes at the cost of positivity:
the most negative eigenvalue of \(\rho_{\text{pred}}\) along the
trajectory approaches \(\approx - 0.10\) across
\(\gamma/J \in \{ 0.1,1,10\}\). The eigenvalue clipping
used during evaluation (\S2.4) keeps fidelity well defined but does not
remove the underlying physical inconsistency; architecturally
constrained completely positive, trace-preserving (CPTP) surrogates
provide the natural path toward resolving this trade-off at the
representation level rather than through post-processing.

Third, the shift of the favorable \(\gamma\) regime, from a bounded
window at \(n = 3\) to a monotonic recovery extending to
\(\gamma/J = 100\) at \(n = 4\), rests on only two nontrivial system
sizes; we do not claim a scaling law relating the optimal dissipation
strength to Hilbert-space dimension. Testing \(n = 5\)(\(d = 32\)) would
begin to test whether the shift persists.

Fourth, the study collapses dephasing and amplitude damping into a
single parameter by setting
\(\gamma_{\phi} = \gamma_{\downarrow} \equiv \gamma\). This keeps the sweep
tractable but prevents disentangling the individual contributions of the
two dissipative mechanisms, which vary across physical NISQ platforms.

Fifth, training uses a Haar-random ensemble while testing uses a
stratified mix (80\% Haar, 10\% computational-basis, 10\% Bell-like;
\S2.3), so the reported fidelities reflect predominantly Haar
performance; we do not claim the mixture bounds Haar-only fidelity
tightly in either direction.

Finally, the closed-system shortfalls
(\(\bar{F}_{\text{avg}} \approx 0.50\) at \(n = 3\) and
\(\approx 0.29\) at \(n = 4\)) are properties of the present architecture
and training budget. We do not claim that closed-system dynamics at
these sizes are fundamentally unlearnable.

\subsection{Implications and outlook}

The principal practical implication is that dissipation can act as a
learning aid by contracting rollout errors. High fidelity at strong
damping does not, however, imply that the surrogate has learned richer
dynamics; it may instead reflect the increasing dominance of the steady
state. The value of learned surrogates therefore lies primarily in the
weak-to-moderate dissipation regime, where they outperform trivial
predictors while still capturing nontrivial transient dynamics.

More broadly, the common intuition that dissipation necessarily
complicates learning is incomplete: for the surrogate class studied
here, the contractive action of dissipation reduces accumulated rollout
error and can enable training objectives, such as the fidelity-aware
loss, that are ineffective in the closed-system (\(\gamma = 0\)) regime
we examined. This connects to the broader Neural ODE literature on
contractive vector fields, where contractivity is known to aid the
stability and generalization of learned dynamical
models.\cite{massaroli2020,revay2024}

Three directions are particularly important. First, and most pressing,
CPTP-constrained architectures (Gram--Hadamard density operators,
purification, or Stinespring-dilation forms) are needed to resolve the
fidelity-versus-positivity trade-off identified in \S5.4; the open
question is whether they cost training stability or
expressiveness.\cite{vicentini2022,torlai2018purification} Second, architecture-independence
studies, in particular Transformer-based trajectory surrogates trained
on the same dataset, would determine whether the \(n = 4\) collapse and
its dissipative rescue are specific to Neural ODEs or reflect a broader
property of learned state-space surrogates. Third, extending the study
to \(n = 5\) and beyond is needed to determine whether the shift in the
favorable dissipation regime persists systematically with Hilbert-space
dimension.

More generally, these results suggest that benchmarking open-system
surrogates with a single aggregate fidelity can obscure the regimes in
which a method genuinely succeeds or fails. The controlled
\(\left( n,\gamma,D \right)\) sweep used here provides a more informative
framework for characterizing the practical reach of learned dynamical
surrogates.

\section{Data and code availability}

For peer review, the complete code and aggregated results accompany this
submission as a supplementary archive. The archive contains the
trajectory-generation code, the Neural ODE training pipeline and
loss-function implementations, all analysis and figure-generation
scripts, the per-run configuration files (one YAML per cell), the
per-cell summary files for all 135 Phase 1 cells and the 18
fidelity-aware extension surrogates, the trivial-predictor baseline data underlying Table~\ref{tab:recovery}, and the per-epoch training logs; all results in the
paper can be regenerated from these files using the included scripts.
Upon acceptance, the same code and data will be released publicly under
an MIT license, and the repository DOI will be included in the published
version.

\section{Author contributions}

A.A. conceived the study, designed the experiments, implemented the
surrogate-training pipeline and data-generation code, performed the
numerical simulations on the University of Michigan Great Lakes cluster,
analyzed the results, and wrote the manuscript.

\section{Competing interests}

The author declares no competing interests.

\section{Acknowledgments}

This work was supported in part by the University of Michigan Office of
the Vice President for Research through a Bold Challenges Booster Award,
and by the
National Science Foundation through the University of Michigan Materials
Research Science and Engineering Center (MRSEC) under Award No.
DMR-2309029. This research was supported in part through computational
resources and services provided by Advanced Research Computing at the
University of Michigan, Ann Arbor.

\bibliographystyle{quantum}
\bibliography{OQSS_Quantum}

\begin{thebibliography}{10}

\bibitem{gorini1976}
V.~Gorini, A.~Kossakowski, and E.~C.~G. Sudarshan.
\newblock ``Completely positive dynamical semigroups of n-level systems''.
\newblock \href{https://dx.doi.org/10.1063/1.522979}{J. Math. Phys. {\bf 17},
  821--825}~(1976).

\bibitem{lindblad1976}
G.~Lindblad.
\newblock ``On the generators of quantum dynamical semigroups''.
\newblock \href{https://dx.doi.org/10.1007/BF01608499}{Commun. Math. Phys. {\bf
  48}, 119--130}~(1976).

\bibitem{breuer2007}
H.~P. Breuer and F.~Petruccione.
\newblock ``The theory of open quantum systems''.
\newblock
  \href{https://dx.doi.org/10.1093/acprof:oso/9780199213900.001.0001}{Oxford
  University Press}. ~(2007).

\bibitem{manzano2020}
D.~Manzano.
\newblock ``A short introduction to the lindblad master equation''.
\newblock \href{https://dx.doi.org/10.1063/1.5115323}{AIP Adv. {\bf 10},
  025106}~(2020).

\bibitem{schollwock2011}
U.~Schollw\"{o}ck.
\newblock ``The density-matrix renormalization group in the age of matrix
  product states''.
\newblock \href{https://dx.doi.org/10.1016/j.aop.2010.09.012}{Ann. Phys. (N.
  Y.) {\bf 326}, 96--192}~(2011).

\bibitem{verstraete2004}
F.~Verstraete, J.~J. Garc\'{i}a-Ripoll, and J.~I. Cirac.
\newblock ``Matrix product density operators: Simulation of finite-temperature
  and dissipative systems''.
\newblock \href{https://dx.doi.org/10.1103/PhysRevLett.93.207204}{Phys. Rev.
  Lett. {\bf 93}, 207204}~(2004).

\bibitem{chen2018node}
R.~T.~Q. Chen, Y.~Rubanova, J.~Bettencourt, and D.~Duvenaud.
\newblock ``Neural ordinary differential equations''.
\newblock Adv. Neural Inf. Process. Syst. (NeurIPS){\bf 31}~(2018).
\newblock  \href{http://arxiv.org/abs/1806.07366}{arXiv:1806.07366}.

\bibitem{kidger2022}
P.~Kidger.
\newblock ``On neural differential equations''.
\newblock PhD thesis.
\newblock University of Oxford.
\newblock ~(2022).

\bibitem{rubanova2019}
Y.~Rubanova, R.~T.~Q. Chen, and D.~K. Duvenaud.
\newblock ``Latent ordinary differential equations for irregularly-sampled time
  series''.
\newblock Adv. Neural Inf. Process. Syst.{\bf 32}~(2019).
\newblock  \href{http://arxiv.org/abs/1907.03907}{arXiv:1907.03907}.

\bibitem{choi2022}
M.~Choi, D.~Flam-Shepherd, T.~H. Kyaw, and A.~Aspuru-Guzik.
\newblock ``Learning quantum dynamics with latent neural ordinary differential
  equations''.
\newblock \href{https://dx.doi.org/10.1103/PhysRevA.105.042403}{Phys. Rev. A
  {\bf 105}, 042403}~(2022).

\bibitem{chen2022}
L.~Chen and Y.~Wu.
\newblock ``Learning quantum dissipation by the neural ordinary differential
  equation''.
\newblock \href{https://dx.doi.org/10.1103/PhysRevA.106.022201}{Phys. Rev. A
  {\bf 106}, 022201}~(2022).

\bibitem{carleo2017}
G.~Carleo and M.~Troyer.
\newblock ``Solving the quantum many-body problem with artificial neural
  networks''.
\newblock \href{https://dx.doi.org/10.1126/science.aag2302}{Science {\bf 355},
  602--606}~(2017).

\bibitem{torlai2018tomography}
G.~Torlai, G.~Mazzola, J.~Carrasquilla, M.~Troyer, R.~Melko, and G.~Carleo.
\newblock ``Neural-network quantum state tomography''.
\newblock \href{https://dx.doi.org/10.1038/s41567-018-0048-5}{Nat. Phys. {\bf
  14}, 447--450}~(2018).

\bibitem{carleo2019netket}
G.~Carleo, K.~Choo, D.~Hofmann, J.~E.~T. Smith, T.~Westerhout, F.~Alet, E.~J.
  Davis, S.~Efthymiou, I.~Glasser, S.~H. Lin, M.~Mauri, G.~Mazzola, C.~B.
  Mendl, E.~van Nieuwenburg, O.~O'Reilly, H.~Th\'{e}veniaut, G.~Torlai,
  F.~Vicentini, and A.~Wietek.
\newblock ``Netket: A machine learning toolkit for many-body quantum systems''.
\newblock \href{https://dx.doi.org/10.1016/j.softx.2019.100311}{SoftwareX {\bf
  10}, 100311}~(2019).

\bibitem{hartmann2019}
M.~J. Hartmann and G.~Carleo.
\newblock ``Neural-network approach to dissipative quantum many-body
  dynamics''.
\newblock \href{https://dx.doi.org/10.1103/PhysRevLett.122.250502}{Phys. Rev.
  Lett. {\bf 122}, 250502}~(2019).

\bibitem{nagy2019}
A.~Nagy and V.~Savona.
\newblock ``Variational quantum monte carlo method with a neural-network ansatz
  for open quantum systems''.
\newblock \href{https://dx.doi.org/10.1103/PhysRevLett.122.250501}{Phys. Rev.
  Lett. {\bf 122}, 250501}~(2019).

\bibitem{vicentini2019}
F.~Vicentini, A.~Biella, N.~Regnault, and C.~Ciuti.
\newblock ``Variational neural-network ansatz for steady states in open quantum
  systems''.
\newblock \href{https://dx.doi.org/10.1103/PhysRevLett.122.250503}{Phys. Rev.
  Lett. {\bf 122}, 250503}~(2019).

\bibitem{yoshioka2019}
N.~Yoshioka and R.~Hamazaki.
\newblock ``Constructing neural stationary states for open quantum many-body
  systems''.
\newblock \href{https://dx.doi.org/10.1103/PhysRevB.99.214306}{Phys. Rev. B
  {\bf 99}, 214306}~(2019).

\bibitem{reh2021}
M.~Reh, M.~Schmitt, and M.~G\"{a}rttner.
\newblock ``Time-dependent variational principle for open quantum systems with
  artificial neural networks''.
\newblock \href{https://dx.doi.org/10.1103/PhysRevLett.127.230501}{Phys. Rev.
  Lett. {\bf 127}, 230501}~(2021).

\bibitem{preskill2018}
J.~Preskill.
\newblock ``Quantum computing in the nisq era and beyond''.
\newblock \href{https://dx.doi.org/10.22331/q-2018-08-06-79}{Quantum {\bf 2},
  79}~(2018).

\bibitem{bharti2022}
K.~Bharti, A.~Cervera-Lierta, T.~H. Kyaw, T.~Haug, S.~Alperin-Lea, A.~Anand,
  M.~Degroote, H.~Heimonen, J.~S. Kottmann, T.~Menke, W.~K. Mok, S.~Sim, L.~C.
  Kwek, and A.~Aspuru-Guzik.
\newblock ``Noisy intermediate-scale quantum algorithms''.
\newblock \href{https://dx.doi.org/10.1103/RevModPhys.94.015004}{Rev. Mod.
  Phys. {\bf 94}, 015004}~(2022).

\bibitem{zakwan2023}
M.~Zakwan, L.~Xu, and G.~Ferrari-Trecate.
\newblock ``Robust classification using contractive hamiltonian neural odes''.
\newblock \href{https://dx.doi.org/10.1109/LCSYS.2022.3186959}{IEEE Control
  Syst. Lett. {\bf 7}, 145--150}~(2023).

\bibitem{qutip1}
J.~R. Johansson, P.~D. Nation, and F.~Nori.
\newblock ``Qutip: An open-source python framework for the dynamics of open
  quantum systems''.
\newblock \href{https://dx.doi.org/10.1016/j.cpc.2012.02.021}{Comput. Phys.
  Commun. {\bf 183}, 1760--1772}~(2012).

\bibitem{qutip2}
J.~R. Johansson, P.~D. Nation, and F.~Nori.
\newblock ``Qutip 2: A python framework for the dynamics of open quantum
  systems''.
\newblock \href{https://dx.doi.org/10.1016/j.cpc.2012.11.019}{Comput. Phys.
  Commun. {\bf 184}, 1234--1240}~(2013).

\bibitem{qutip5}
N.~Lambert, E.~Gigu\`{e}re, P.~Menczel, B.~Li, P.~Hopf, G.~Su\'{a}rez, M.~Gali,
  J.~Lishman, R.~Gadhvi, R.~Agarwal, A.~Galicia, N.~Shammah, P.~Nation, J.~R.
  Johansson, S.~Ahmed, S.~Cross, A.~Pitchford, and F.~Nori.
\newblock ``Qutip 5: The quantum toolbox in python''.
\newblock \href{https://dx.doi.org/10.1016/j.physrep.2025.10.001}{Phys. Rep.
  {\bf 1153}, 1--62}~(2026).

\bibitem{vicentini2022}
F.~Vicentini, R.~Rossi, and G.~Carleo.
\newblock ``Positive-definite parametrization of mixed quantum states with deep
  neural networks''~(2022).
\newblock  \href{http://arxiv.org/abs/2206.13488}{arXiv:2206.13488}.

\bibitem{mikeska2004}
H.-J. Mikeska and A.~K. Kolezhuk.
\newblock ``One-dimensional magnetism''.
\newblock In U.~Schollw\"{o}ck, J.~Richter, D.~J.~J. Farnell, and R.~F. Bishop,
  editors, Quantum Magnetism.
\newblock \href{https://dx.doi.org/10.1007/BFb0119591}{Volume 645 of Lecture
  Notes in Physics, pages 1--83}.
\newblock Springer, Berlin, Heidelberg~(2004).

\bibitem{uhlmann1976}
A.~Uhlmann.
\newblock ``The ``transition probability'' in the state space of a
  $*$-algebra''.
\newblock \href{https://dx.doi.org/10.1016/0034-4877(76)90060-4}{Rep. Math.
  Phys. {\bf 9}, 273--279}~(1976).

\bibitem{jozsa1994}
R.~Jozsa.
\newblock ``Fidelity for mixed quantum states''.
\newblock \href{https://dx.doi.org/10.1080/09500349414552171}{J. Mod. Opt. {\bf
  41}, 2315--2323}~(1994).

\bibitem{nielsen2010}
M.~A. Nielsen and I.~L. Chuang.
\newblock ``Quantum computation and quantum information: 10th anniversary
  edition''.
\newblock \href{https://dx.doi.org/10.1017/CBO9780511976667}{Cambridge
  University Press}. ~(2010).

\bibitem{smolin2012}
J.~A. Smolin, J.~M. Gambetta, and G.~Smith.
\newblock ``Efficient method for computing the maximum-likelihood quantum state
  from measurements with additive gaussian noise''.
\newblock \href{https://dx.doi.org/10.1103/PhysRevLett.108.070502}{Phys. Rev.
  Lett. {\bf 108}, 070502}~(2012).

\bibitem{dormand1980}
J.~R. Dormand and P.~J. Prince.
\newblock ``A family of embedded runge-kutta formulae''.
\newblock \href{https://dx.doi.org/10.1016/0771-050X(80)90013-3}{J. Comput.
  Appl. Math. {\bf 6}, 19--26}~(1980).

\bibitem{elfwing2018}
S.~Elfwing, E.~Uchibe, and K.~Doya.
\newblock ``Sigmoid-weighted linear units for neural network function
  approximation in reinforcement learning''.
\newblock \href{https://dx.doi.org/10.1016/j.neunet.2017.12.012}{Neural Netw.
  {\bf 107}, 3--11}~(2018).

\bibitem{ramachandran2017}
P.~Ramachandran, B.~Zoph, and Q.~V. Le.
\newblock ``Searching for activation functions''.
\newblock 6th International Conference on Learning Representations (ICLR),
  Workshop Track~(2017).
\newblock  \href{http://arxiv.org/abs/1710.05941}{arXiv:1710.05941}.

\bibitem{kingma2014}
D.~P. Kingma and J.~L. Ba.
\newblock ``Adam: A method for stochastic optimization''.
\newblock 3rd International Conference on Learning Representations
  (ICLR)~(2014).
\newblock  \href{http://arxiv.org/abs/1412.6980}{arXiv:1412.6980}.

\bibitem{massaroli2020}
S.~Massaroli, M.~Poli, J.~Park, A.~Yamashita, and H.~Asama.
\newblock ``Dissecting neural odes''.
\newblock Adv. Neural Inf. Process. Syst. {\bf 33}, 3952--3963~(2020).
\newblock  \href{http://arxiv.org/abs/2002.08071}{arXiv:2002.08071}.

\bibitem{revay2024}
M.~Revay, R.~Wang, and I.~R. Manchester.
\newblock ``Recurrent equilibrium networks: Flexible dynamic models with
  guaranteed stability and robustness''.
\newblock \href{https://dx.doi.org/10.1109/TAC.2023.3294101}{IEEE Trans.
  Automat. Contr. {\bf 69}, 2855--2870}~(2024).

\bibitem{torlai2018purification}
G.~Torlai and R.~G. Melko.
\newblock ``Latent space purification via neural density operators''.
\newblock \href{https://dx.doi.org/10.1103/PhysRevLett.120.240503}{Phys. Rev.
  Lett. {\bf 120}, 240503}~(2018).

\end{thebibliography}

\end{document}